\documentclass[PRL,preprint,superscriptaddress,showpacs,preprintnumbers,amsmath,amssymb,floats]{revtex4}
\usepackage{graphicx}

\usepackage[normalem]{ulem}

\begin{document}

\title{High-resolution inelastic X-ray scattering studies of the state-resolved differential cross-section of Compton excitations in helium
atoms}
\author{B.~P.~Xie}
\affiliation{Department of Physics, Surface Physics Laboratory
(National Key Laboratory), and Advanced Materials Laboratory, Fudan
University, Shanghai 200433, P. R. China}
\author{L.~F.~Zhu}
\email{lfzhu@ustc.edu.cn}\affiliation{Hefei National Laboratory for
Physical Sciences at Microscale, Department of Modern Physics,
University of Science and Technology of China, Hefei, Anhui, 230026,
P. R. China}
\author{K.~Yang}  \altaffiliation[Present affiliation: ]{Shanghai institute of applied physics, Chinese Academy of Sciences. P. R. China.}
\author{B.~Zhou}
\affiliation{Department of Physics, Surface Physics Laboratory
(National Key Laboratory), and Advanced Materials Laboratory, Fudan
University, Shanghai 200433, P. R. China}
\author{N.~Hiraoka}
\author{Y.~Q.~Cai}
\altaffiliation[Present affiliation: ]{NSLS-II,
Brookhaven National Laboratory, USA.} \affiliation{National
Synchrotron Radiation Research Center, Hsinchu 30076, Taiwan,
Republic of China}
\author{Y.~Yao}
\author{C.~Q.~Wu}
\affiliation{Department of Physics, Fudan University, Shanghai
200433, P. R. China}
\author{E.~L.~Wang}
\affiliation{Hefei National Laboratory for Physical Sciences at
Microscale, Department of Modern Physics, University of Science and
Technology of China, Hefei, Anhui, 230026, P. R. China}
\author{D.~L.~Feng}
\email{dlfeng@fudan.edu.cn} \affiliation{Department of Physics,
Surface Physics Laboratory (National Key Laboratory), and Advanced
Materials Laboratory, Fudan University, Shanghai 200433, P. R.
China}

\date{\today}

\begin{abstract}

The state-resolved differential cross sections for both the $1s^2$
$^1S_0 \rightarrow 1s2s$ $^1S_0$ monopolar transition  and the
$1s^2$ $^1S_0 \rightarrow 1s2p$ $^1P_1$ dipolar transition of atomic
helium have been measured over a large momentum transfer region by
the high-resolution inelastic X-ray scattering (IXS) for the first
time. The almost perfect match of the present measurement with the
theoretical calculations gives a stringent test of the theoretical
method and the calculated wavefunctions. Our results demonstrate
that high-resolution IXS is a powerful tool for studying the
excitations in atoms and molecules.

\end{abstract}

\pacs{34.50.Fa, 32.30.Rj, 32.70.Cs} \maketitle

After the discovery of Compton effect in 1923, it was soon
recognized that Compton effect can give valuable information about
the electronic momentum density of the target~\cite{DuMond}. About
forty years later, Platzman and coworkers found that the Compton
profile measured by the cross section of the Compton ionization,
gives information on the electron momentum distribution of the
ground state; while the Compton excitation measured through
inelastic X-ray scattering (IXS) or X-ray Raman scattering, can be
used to probe the wavefunctions of electrons in excited state
\cite{Platzman}. The differential cross section (DCS) of Compton
excitation can be described by:
$$\frac{d^2\sigma}{d\Omega d\omega_f}= r_0^2 \frac{\omega_f}{\omega_i}
|\vec{\epsilon_i} \cdot \vec{\epsilon_f^*} |^2 S(\vec{q},\omega),
$$ where  $S(\vec{q},\omega)$ is the dynamical structure factor defined as
$$S(\vec{q},\omega)= \sum_{f} |\langle \psi_f |  e^{-i
\vec{q} \cdot \vec{r}} | \psi_i \rangle|^{2}
\delta(E_i-E_f+\hbar(\omega_i-\omega_f)),$$
$\omega=\omega_i-\omega_f$ is the energy loss, and
$\vec{q}=\vec{k_i}-\vec{k_f}$ is the scattering vector.
$S(\vec{q},\omega)$ provides a more comprehensive description of the
excitations than usual optical measurements, such as
photo-absorption, since it reveals the momentum distribution
character related to the initial and final state wavefunctions
($|\psi_i\rangle$ and $|\psi_f\rangle$)~\cite{amusia}. The
determination of the wavefunctions of a quantum many-body system,
even as simple as a helium atom, is a challenging task, particularly
for the excited states. Therefore, the study of the Compton
excitation would be of great importance in building the fundamental
pictures of quantum mechanics, atoms, molecules and solids.

Although IXS has been known for over forty years, this technique was
mostly adopted to the study of condensed matter systems with high
density, due to the very small cross section of IXS~\cite{cai}. For
example, IXS has been proved successful in studying charge
excitations of superconductors~\cite{caiMgB2},
graphite~\cite{Wendymao}, ice~\cite{caiice}, and organic molecular
crystals~\cite{KYang}. The measurement of Compton excitation on
low-density gas phase subject is hampered by the very weak signal
that cannot fulfill the high energy resolution requirement
($E/\Delta E \sim 10^{5}$) of the state-resolved measurement. In
recent years, the advance of synchrotron techniques has enabled
three pieces of investigation, to the best of our knowledge, on
Compton excitations in gases. \v{Z}itnik \emph{et al.} reported the
Compton excitation spectra in the vicinity of xenon $L_3$ edge at a
scattering angle of 90$^\circ$ with an energy resolution of 1\,eV
\cite{zitnik}. Kav\v{c}i\v{c} \emph{et al.} reported the spectra of
argon gas, where the features related to different two-electron
atomic processes in the vicinity of an inner-shell threshold are
separated with an energy resolution of 0.6 eV~\cite{kavcic}.
Moreover, the Compton excitation spectra for the doubly excited
states of helium at three angles were reported by Minzer \emph{et
al.} with an energy resolution of 0.9\,eV~\cite{minzer}.  However,
an energy resolutions of 0.6$\sim$1 eV of these studies are not
enough to resolve the adjacent transitions clearly, especially for
excitations of valence shell state in atoms and molecules.
Furthermore, there is so far no measurement of the DCS or dynamic
structure factor $S(\vec{q},\omega)$ of Compton excitation for gas
phase subject.

In this letter, as the first attempt to measure the state-resolved
differential cross sections of Compton excitation for gas phase, we
investigated the $S(\vec{q},\omega)$ of the $1s^2$ $^1S_0
\rightarrow 1s2s$ $^1S_0$  and  $1s^2$ $^1S_0 \rightarrow 1s2p$
$^1P_1$ transitions of atomic helium with ultra high energy
resolution and over a large momentum transfer region by IXS. Spectra
have been measured with an energy resolution as high as 70~meV. The
selection of helium is because it is the simplest multi-electron
system, for which  a reliable theoretical calculation can be
achieved. Therefore,  the $S(\vec{q},\omega)$ measured with a high
resolution would provide a benchmark  to test the theoretical method
stringently~\cite{han,Cann}. Historically, helium played an
important role in the development of Compton profile since a
meaningful comparison between theory and experiment was feasible
only on helium atom at that time
\cite{Eisenberger1,Eisenberger2,Wellenstein}. Furthermore, the DCS's
for $1s^2$ $^1S_0 \rightarrow 1s2s$ $^1S_0$ and $1s^2$ $^1S_0
\rightarrow 1s2p$ $^1P_1$ transitions have been studied by the high
energy electron-energy-loss spectroscopy (EELS)~\cite{LiuJES,Xu},
and the present measurement by IXS provides an independent
cross-check.

\begin{figure}[t!]
\includegraphics[width=7.5cm]{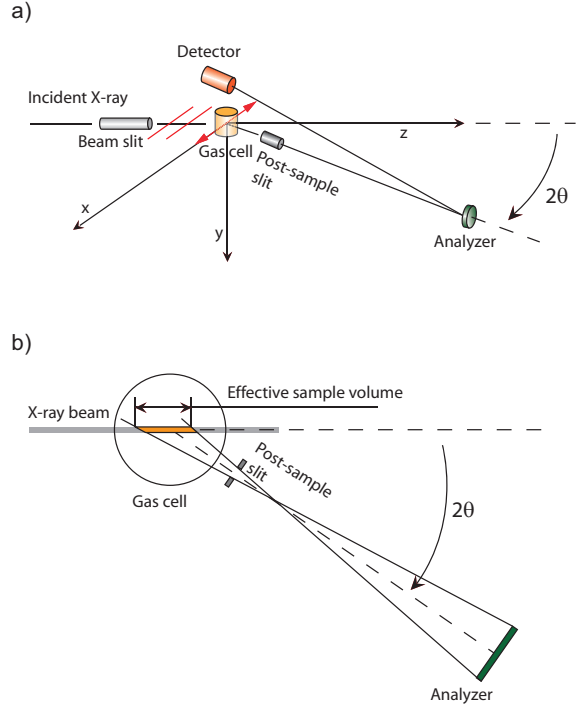}
\caption{(color online) (a) Schematics of the experimental setup for
inelastic X-ray scattering of gases. The polarization direction of
incident photon is along x axis. (b) Deduction of the effective
sample volume (top view).}\label{setup}
\end{figure}


\begin{figure}[t!]
\includegraphics[width=7.5cm]{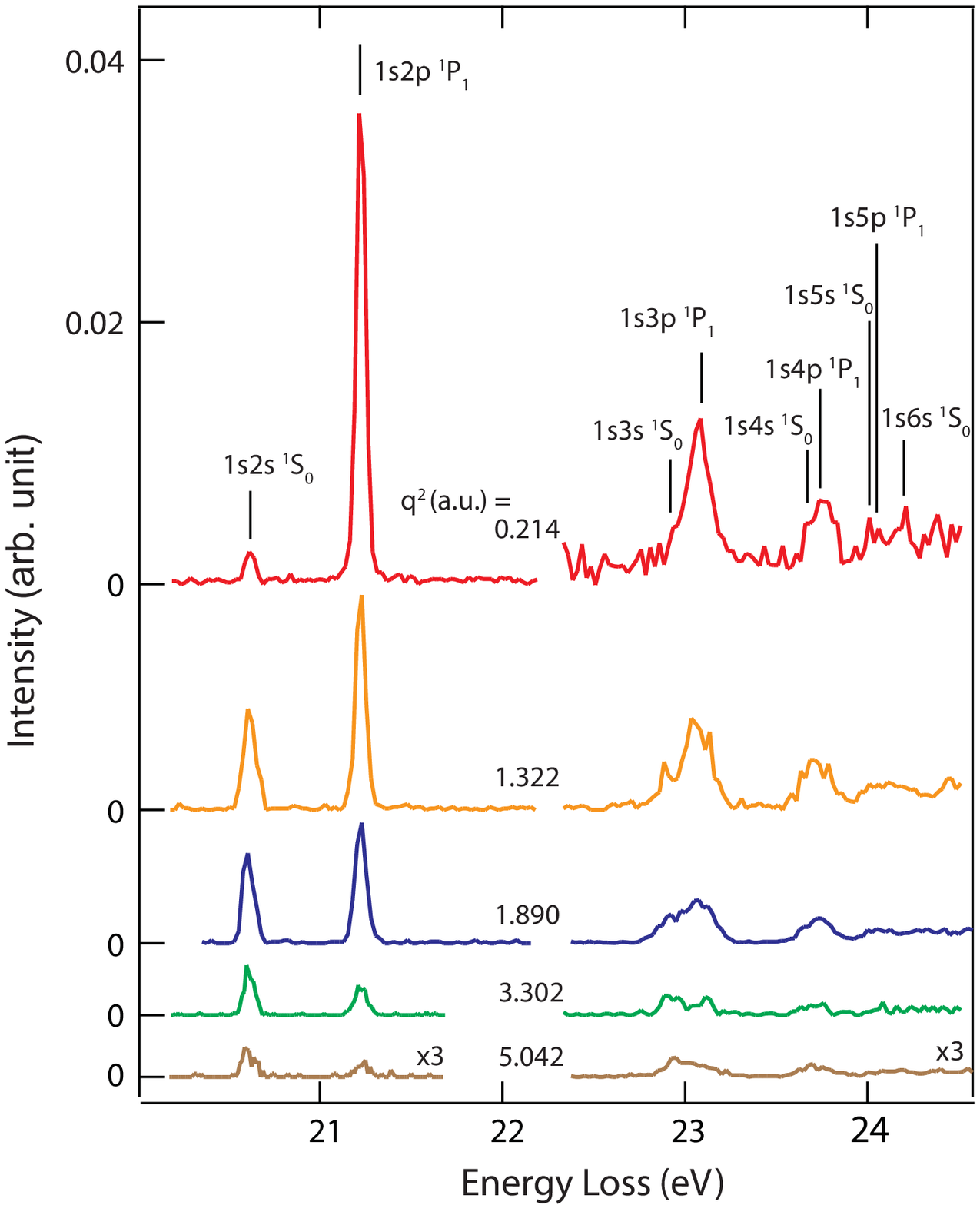}
\caption{(color online) Inelastic X-ray scattering spectra of helium
gas as a function of energy loss at different momentum transfers,
and peaks are labeled by the final state configurations excited from
the $1s^2$ $^1S_0$ ground state. The spectra at lower energy loss
range were collected with a resolution of 70~meV, while  those at
higher energy loss range were gathered with a resolution of
170~meV.}\label{spectra}
\end{figure}

The present IXS measurements were carried out at the Taiwan Beamline
BL12XU of SPring-8 \cite{cai}. 3\,atm of helium gas was sealed in a
gas cell with Kapton windows [Fig.~\ref{setup}(a)]. The energy
spread of the incoming beam depends on the monochromator. Two energy
resolution setups were exploited. A Si(333) high-resolution
monochromator or a Si(400) monochromator is in place to achieve the
resolution of 70 meV or 170 meV respectively. Si(555) spherical
analyzers with 2 m radius of curvature were used to collect the
scattered photons. The analyzer energy for the scattered photon was
fixed at 9889.68 eV, while the incident photon energy varied, from
which the energy loss is deduced. The momentum resolution is about
0.17$\AA^{-1}$ (or 0.091 a.u. in atomic units). The total absorption
of the X-ray is about 13\%, which mostly comes from the Kapton
window and is constant in the scanned energy range. The scattering
signal from the Kapton window is far from the center of gas cell,
and is mostly blocked by the post-sample slit. It does contribute to
a tiny constant background at small $2\theta$ angle region, which
can be easily removed during data analysis. The incident beam spot
is about $80\times 120$ $\mu m^2$. All of the data were taken at
room temperature. As illustrated in Fig.~\ref{setup}(b),
 the different lengths of
the scattering pathway would be detected with varying $2\theta$,
because of the finite size of the gas cell. This leads to an angular
factor of $\sin(2\theta)$, which needs to be corrected to extract
the DCS of Compton excitation. The fluctuation of the incident X-ray
intensity is monitored through a Silicon PIN diode and can be
corrected accordingly.

Selected IXS spectra of helium are shown in Fig.~\ref{spectra}, and
the assignments for the transitions are based on the energy
positions of the NIST database~\cite{NIST}. Different from the
photo-absorption dominated by the dipolar transition, IXS can excite
not only the dipolar transitions but also the transitions of other
multipolarity~\cite{Platzman,amusia,Inokuti}. As a result, the
$1s^2$ $^1S_0 \rightarrow 1s2s$ $^1S_0$ monopolar transition
\cite{mono} and the $1s^2$ $^1S_0 \rightarrow 1s2p$ $^1P_1$ dipolar
transition are clearly resolved at the present energy resolution of
70 meV, which is the best energy resolution achieved for the gas
target of IXS and even slightly better than 80 meV of  the high
energy EELS~\cite{liu,LiuJES}. The measured peak width is limited by
the energy resolution. The transitions at higher energy loss range
are not clearly resolved due to the limited energy resolution of 170
meV. More detailed infromation is expected to be revealed with a
higher energy resolution in the future. It can also be seen from
Fig.~\ref{spectra} that the monopolar transition is much weaker than
the dipolar transition at low momentum transfer, while the situation
is reversed at high momentum transfer. This illustrates that the
dipole approximation for IXS cross section breaks down in the high
momentum transfer region~\cite{IXSreview}. This was reported
previously in lithium metal \cite{IXSLi}, but the current results
give a much clearer exemplification of this basic fact of IXS.
Similar situation has been observed in high energy EELS experiments
before~\cite{Chan1991}.

\begin{figure}[t!]
\includegraphics[width=7.5cm]{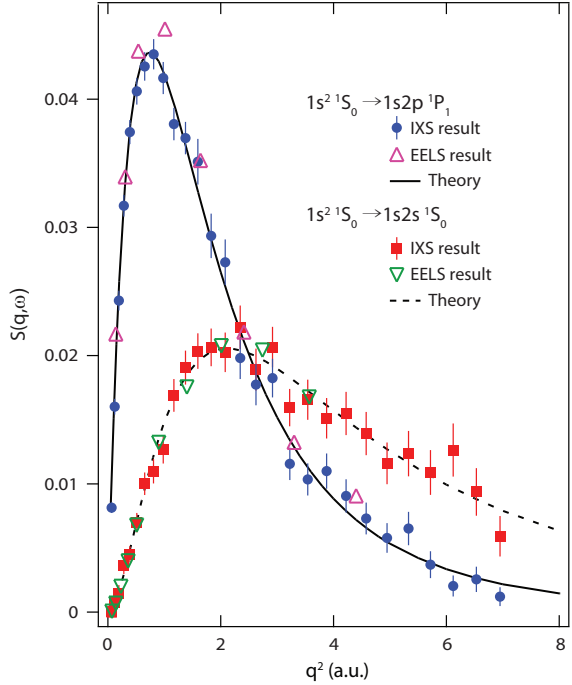}
\caption{(color online) The dynamic structure factor
$S(\vec{q},\omega)$ of the $1s^2$ $^1S_0 \rightarrow 1s2s$ $^1S_0$
monopolar transition and the $1s^2$ $^1S_0 \rightarrow 1s2p$ $^1P_1$
dipolar transition of helium measured by IXS. The high energy EELS
data~\cite{LiuJES,Xu} and the theoretical calculations~\cite{Cann}
were converted from the generalized oscillator strengths. Data were
taken with 170\,meV resolution. Note the horizontal axis is
$q^2=k_i^2+k_f^2-2k_ik_f\cos(2\theta)$, following the convention in
atomic physics.}\label{qdep}
\end{figure}

To determine the DCS's of the $1s^2$ $^1S_0 \rightarrow 1s2s$
$^1S_0$ and $1s^2$ $^1S_0 \rightarrow 1s2p$ $^1P_1$ excitations,
their respective scattered photon intensities at the  energy loss
positions of 20.616 eV and 21.218 eV  were measured by scanning the
scattering angle $2\theta$ from $5^\circ$ to $60^\circ$. To
calibrate the background,  a $2\theta$ scan was run at 20.1\,eV
energy loss, since it is in a flat and featureless region of the IXS
spectrum [Fig.~\ref{spectra}]. After correcting the contribution of
background and effective scattering volume at different scattering
angles [Fig.~\ref{setup}(b)], the dynamic structure factor
$S(\vec{q},\omega)$ for these two transitions were obtained via
dividing the DCS's by the factor of
$\cos^2(2\theta)\omega_f/\omega_i$ for the incident photon
polarization in the scattering plane, and scaling the value of
$1s^2$ $^1S_0 \rightarrow 1s2p$ $^1P_1$ transition to the
theoretical one at $q^2=0.81$ a.u. Fig.~\ref{qdep} shows the
measured dynamic structure factor $S(\vec{q},\omega)$ of the
monopolar excitation of $1s^2$ $^1S_0 \rightarrow 1s2s$ $^1S_0$ and
the dipolar one of $1s^2$ $^1S_0 \rightarrow 1s2p$ $^1P_1$ along
with some previous high energy EELS data~\cite{LiuJES,Xu}.
Theoretical calculations based on the explicitly correlated wave
functions are also presented~\cite{Cann}. Compared with the commonly
used Hylleraas-type wave functions, Cann and Thakkar used the
exponential correlation factors to describe the electron
correlations in their calculation, and the calculated energies are
in agreement with the best published values within
nanohartrees~\cite{Cann}. It can be seen from Fig.~\ref{qdep} that
the almost perfect match of the theoretical curves with the present
data in a broader momentum range confirms that the calculated wave
functions is close to the actual wave functions to a great detail in
real space. It can also be noticed that the $S(\vec{q},\omega)$ for
the dipole excitation is concentrated in the low momentum transfer,
while that for the monopole excitation is much broader in momentum
space. This can be reasonably explained by the fact that the wave
function of $1s2p$ $^1P_1$ is more extended in real space than that
of $1s2s$ $^1S_0$.

It is worth noting that IXS and high energy EELS are complementary.
Measuring $S(\vec{q},\omega)$ by these two independent techniques
provides a means of data cross-check. Because the extraction of
$S(\vec{q},\omega)$ from high energy EELS data is based on the
validity of the first Born approximation (FBA), which requires a
high incident electron energy~\cite{Inokuti}.  As for how high the
incident energy has to be, it can only be determined on a
trial-and-error basis. However, IXS is more
straightforward~\cite{Platzman}, so that its cross section can serve
as a benchmark to test whether the FBA holds in EELS experiments.
Moreover, high energy EELS has much larger cross sections than IXS
for the lower momentum transfer. However, this cross-section
advantage of the electron scattering method rapidly diminishes with
a rate of $q^{-4}$ (Rutherford cross section) as the momentum
transfer increases, while the cross section of Compton scattering is
simply proportional to $S(\vec{q},\omega)$. It is also known that
high energy EELS suffers from multiple scattering at high $q$, while
the IXS does not. Generally speaking, high energy EELS is more
effective to study the large-scale structures of wavefunctions
(lower $q$), while IXS is reliable at all length scales. As a
result, the collection of IXS data presented in Fig.~\ref{qdep} over
such a broad momentum range only takes several hours, while the high
energy EELS data requires several weeks. Since the pressure in the
gas cell could be further increased, there is a tremendous potential
for IXS in studying the excitations of atoms and molecules.

Practically,  IXS could be readily extended to more applied research
where the gas systems may be subject to various extreme physical and
chemical conditions, such as in catalysis, and under extreme
pressure and temperature conditions. The IXS data may be used, for
example, to identify chemical species \textit{in situ} as they are
produced. This type of experimental environments is beyond the reach
of other experimental techniques such as EELS and photoemission.

In conclusion, we have demonstrated that IXS is a powerful tool to
study the excitations in atomic or molecular systems at a third
generation synchrotron. The dynamic structure factors of  the $1s^2$
$^1S_0 \rightarrow 1s2s$ $^1S_0$ monopolar excitation and the $1s^2$
$^1S_0 \rightarrow 1s2p$ $^1P_1$  dipolar excitation of atomic
helium have been measured by the high-resolution IXS over a wide
momentum transfer range.  The almost perfect match of the present
measurement with the theoretical calculations gives a rigorous test
of the theoretical method, and demonstrates the cleanness of the
data. Furthermore, our data provide a benchmark for the direct
determination of the absolute DCS values of other gases in future
experiments, {\sl e.g.}, with a mixed-gas setup.

We gratefully acknowledge the helpful discussions with Professor G.
A. Sawatzky, and Professor H. Chen. This work was supported by the
NSFC (Grants No. 20673023, 10634030, 10734040, 10874168),  MOST
(National Basic Research Program No.2006CB921300), CAS (Knowledge
Promotion Project No. KJCS1-YW-N30), and STCSM of China. The
experiment was carried out in a beamtime approved by JSRRI, Japan
(Proposal No. 2008A4262) and NSRRC, Taiwan, Republic of China
(No.2008-1-002-2).

\end{document}